\documentclass[sigconf]{acmart}

\renewcommand\footnotetextcopyrightpermission[1]{} 
\AtBeginDocument{%
  \providecommand\BibTeX{{%
    \normalfont B\kern-0.5em{\scshape i\kern-0.25em b}\kern-0.8em\TeX}}}

\setcopyright{none}

\usepackage{xcolor}

\begin{document}
\fancyhead{}
\title{Detection of Novel Social Bots\\by Ensembles of Specialized Classifiers}




\author{Mohsen Sayyadiharikandeh,$^{1}$*
Onur Varol,$^2$ Kai-Cheng Yang,$^1$ Alessandro Flammini,$^{1}$}\thanks{*Corresponding Author. Email: msayyadi@indiana.edu}

\author{Filippo Menczer$^{1}$}

\affiliation{%
  \institution{\textsuperscript{\rm 1}Observatory on Social Media, Indiana University, Bloomington, IN, USA}
}

\affiliation{%
  \institution{\textsuperscript{\rm 2}Faculty of Engineering and Natural Sciences, Sabanci University, Istanbul, Turkey}
}

\renewcommand{\shortauthors}{Sayyadiharikandeh et al.}

\begin{abstract}
Malicious actors create inauthentic social media accounts controlled in part by algorithms, known as social bots, to disseminate misinformation and agitate online discussion.
While researchers have developed sophisticated methods to detect abuse, novel bots with diverse behaviors evade detection. 
We show that different types of bots are characterized by different behavioral features. As a result, supervised learning techniques suffer severe performance deterioration when attempting to detect behaviors not observed in the training data.
Moreover, tuning these models to recognize novel bots requires retraining with a significant amount of new annotations, which are expensive to obtain.
To address these issues, we propose a new supervised learning method that trains classifiers specialized for each class of bots and combines their decisions through the maximum rule.
The ensemble of specialized classifiers (ESC) can better generalize, leading to an average improvement of 56\% in F1 score for unseen accounts across datasets.
Furthermore, novel bot behaviors are learned with fewer labeled examples during retraining. 
We deployed ESC in the newest version of Botometer, a popular tool to detect social bots in the wild, with a cross-validation AUC of 0.99. 
\end{abstract}



\maketitle
\pagestyle{empty}

\section{Introduction}

Social media accounts partially controlled  by algorithms, known as social bots, have been extensively studied \cite{ferrara2016rise,subrahmanian2016darpa}.
The automated nature of bots makes it easy to achieve scalability when spreading misinformation \cite{shao2018spread,shao2018anatomy}, amplifying popularity \cite{ratkiewicz2011truthy,cresci2015fame,varol2019journalists}, or polarizing online discussion \cite{stella2018bots}.
Bot activity has been reported in different domains, including politics \cite{bessi2016social,ferrara2017disinformation,stella2018bots}, health \cite{allem2017cigarette,allem2018could,deb2018social,broniatowski2018weaponized}, and business \cite{cresci2018fake,cresci2019cashtag}.
Due to the wide adoption of social media, every aspect of people's life from news consumption to elections is vulnerable to potential manipulation by bots.

The public is beginning to recognize the existence and role of social bots: according to a recent Pew survey \cite{stocking2018pew}, two thirds of Americans have knowledge of bots and over 80\%  believe bots have a negative impact.
Actions have been taken to restrict the potential damage caused by deceptive bots, such as those that pose as humans.
For example, California passed a ``Bot Disclosure'' law in July 2019, requiring bots to reveal themselves in certain cases (\url{leginfo.legislature.ca.gov/faces/billTextClient.xhtml?bill_id=201720180SB1001}). 
However, there is no guarantee that such legislative solutions will be effective against malicious bots, or even that they will survive constitutional challenges.
In an arms race between abusive behaviors and countermeasures, novel social bots emerge everyday and evade purge from the platforms \cite{ferrara2016rise,cresci2019better}.
Therefore, the availability of tools to identify social bots is still important for protecting the authenticity and health of the information ecosystem.

Many social bot detection methods based on machine learning have been proposed in the past several years (see Related Work). Here we focus on supervised learning methods, particularly Botometer \cite{varol2017online,yang2019arming}, a widely adopted tool designed to evaluate Twitter accounts.
Supervised methods are only as good as their training data.
Bots with unseen characteristics are easily missed, as demonstrated by a drastic drop in recall when classifiers are faced with cross-domain accounts \cite{de2018lobo}. 
One common approach to address the lack of generalization is to retrain models with new labeled datasets \cite{yang2019arming}. 
Unfortunately, high-quality datasets of annotated social media accounts are expensive to acquire.

In this paper, we aim to improve the cross domain performance of Botometer in the wild and better tune the method to the adversarial bot detection problem. 
Using bot and human accounts in different datasets, we show that bot accounts exhibit greater heterogeneity in their discriminative behavioral features compared to human accounts. 
Motivated by this observation, we propose a novel method to construct a bot detection system capable of better generalization by training multiple classifiers specialized for different types of bots.
Once these domain-specific classifiers are trained, unseen accounts are evaluated by combining their assessments.
We evaluate \emph{cross-domain} performance by testing on datasets that are not used for training, as opposed to \emph{in-domain} evaluation through cross-validation.  
Without loss of in-domain accuracy, the proposed approach effectively increases the recall of cross-domain bots.
It can also learn more efficiently from examples in new domains. 

Given these results, the proposed method is deployed in the newest version (v4) of Botometer, a widely adopted tool to detect social bots in the wild that is publicly available from the Observatory on Social Media at Indiana University. 

\begin{table}[t!]
\caption{Annotated datasets.}
\centering
\small
\begin{tabular}{lp{2.5cm}lrr}
   \hline
   Dataset & Annotation method & Ref. & Bots & Humans\\
   \hline
   \texttt{\footnotesize caverlee}  & Honeypot + verified & \citealp{lee2011seven} & 15,483 & 14,833\\
   \texttt{\footnotesize varol-icwsm} & Human annotation & \citealp{varol2017online} & 733 & 1,495\\
   \texttt{\footnotesize cresci-17} & Various methods & \citealp{cresci2017paradigm} & 7,049 & 2,764\\
   \texttt{\footnotesize pronbots} & Spam bots & \citealp{yang2019arming} & 17,882 & 0\\
   \texttt{\footnotesize celebrity} & Celebrity accounts & \citealp{yang2019arming} & 0 & 5,918\\
   \texttt{\footnotesize vendor-purchased} & Fake followers &\citealp{yang2019arming} & 1,087 & 0\\
   \texttt{\footnotesize botometer-feedback} & Human annotation & \citealp{yang2019arming} & 139 & 380\\
   \texttt{\footnotesize political-bots} & Human annotation & \citealp{yang2019arming} & 62 & 0\\
   \texttt{\footnotesize gilani-17} & Human annotation  & \citealp{gilani2017bots} & 1,090 & 1,413\\
   \texttt{\footnotesize cresci-rtbust} & Human annotation &\citealp{mazza2019rtbust} & 353 & 340\\
   \texttt{\footnotesize cresci-stock} & Sign of coordination & \citealp{cresci2018fake} & 7,102 & 6,174\\
   \texttt{\footnotesize botwiki} & Human annotation  & \citealp{yang2020scalable}  & 698 & 0\\
   \texttt{\footnotesize astroturf} & Human annotation &  & 505  & 0\\
   \texttt{\footnotesize midterm-2018} & Human annotation & \citealp{yang2020scalable} & 0 & 7459\\
   \texttt{\footnotesize kaiser-1} & Politicians + new bots  & \citealp{botometerCriticismHarvard}  & 875 & 499\\
   \texttt{\footnotesize kaiser-2} & German politicians + German bots  & \citealp{botometerCriticismHarvard}  & 27 & 508\\
   \texttt{\footnotesize kaiser-3} & German politicians + new bots & \citealp{botometerCriticismHarvard}  & 875 & 433\\
   \texttt{\footnotesize combined-test} & \texttt{\footnotesize gilani-17} + \texttt{\footnotesize cresci-rtbust} + \texttt{\footnotesize cresci-stock} + \texttt{\footnotesize kaiser-1} + \texttt{\footnotesize kaiser-2} & & 9,432 & 8,862 \\
   \hline
\end{tabular}
   \label{table:dataset} 
\end{table}

\section{The challenge of generalization}

\subsection{Datasets}

We considered various labeled datasets available through the Bot Repository (\url{botometer.iuni.iu.edu/bot-repository}). 
Most of the data\-sets are annotated by humans, while others are created using automated techniques based on account behavior, filters on metadata, or more sophisticated procedures to achieve high precision. 
For example, \texttt{astroturf} is a new dataset that includes hyper-active political bots participating in follow trains and/or systematically deleting content.
Detailed dataset descriptions are outside the scope of the present paper, but summary statistics and references can be found in Table~\ref{table:dataset}. 
In addition to the datasets in the Bot Repository, we also collected accounts provided in a recent study by \citet{botometerCriticismHarvard}. These datasets made an assumption that all accounts belonging to American and German politicians are human accounts. They complement this dataset with manually annotated German language bots and accounts listed in the \texttt{botwiki} dataset.

For training models, we extract over 1,200 features in six categories: metadata from the accounts and friends, retweet/mention networks, temporal features, content information, as well as sentiment. These features are shown to be effective in identifying social bots and are described in detail in the literature \cite{varol2017online,varol2018feature,yang2019arming}.

\subsection{Cross-domain performance comparison}

\begin{figure}[t!]
    \centering
    \includegraphics[width=\columnwidth]{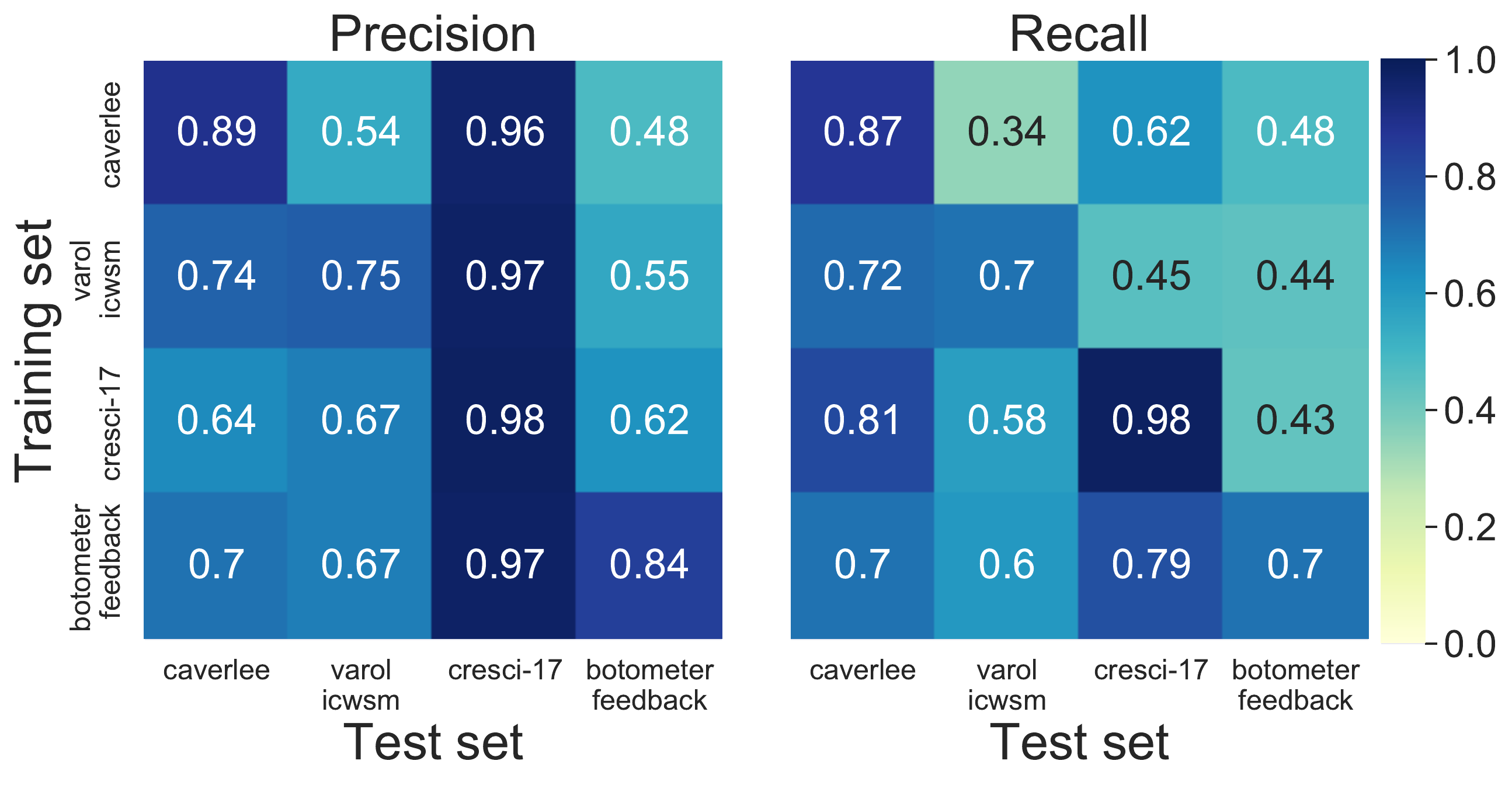}
    \caption{
    Precision (left) and recall (right) of Random Forests trained on one dataset (row) and tested on another (column).
    }
    \label{fig:precRecHeatmap}
\end{figure}

Supervised bot detection methods  achieve high accuracy based on in-domain cross-validation \cite{alothali2018detecting}.
To measure how recall deteriorates in cross-domain evaluation, we perform an experiment using four datasets selected from Table~\ref{table:dataset}: we train a model on one dataset and test it on another. 
We use Random Forest classifiers with 100 decision trees (similar to the baseline model described in \S~\ref{sec:baseline}). The fraction of trees outputting a positive label is calibrated using Platt's scaling~\cite{calibrationPaper} and binarized with a threshold of 0.5. 
We use 5-fold cross-validation for in-domain classification; for consistency we split training and test samples in cross-domain cases as well, reporting average precision and recall. 

The results of our experiment are shown in Fig.~\ref{fig:precRecHeatmap}.
Diagonal (off-diagonal) cells represent in-domain (cross-domain) performance.
Both precision and recall tend to be higher for in-domain cases, demonstrating the limited generalization of supervised models across domains.
The one exception is the high precision when testing on the \texttt{cresci-17} domain, irrespective of the training datasets. This is due to the fact that \texttt{cresci-17} includes spambots, which are represented in all datasets. 
By comparing the two panels, we see that recall of bots is more impaired in cross-domain tests, in line with previous findings \cite{cresci2017paradigm,de2018lobo}.
The method proposed here improves cross-domain bot recall. 

\begin{figure}
    \centering
    \includegraphics[width=\columnwidth]{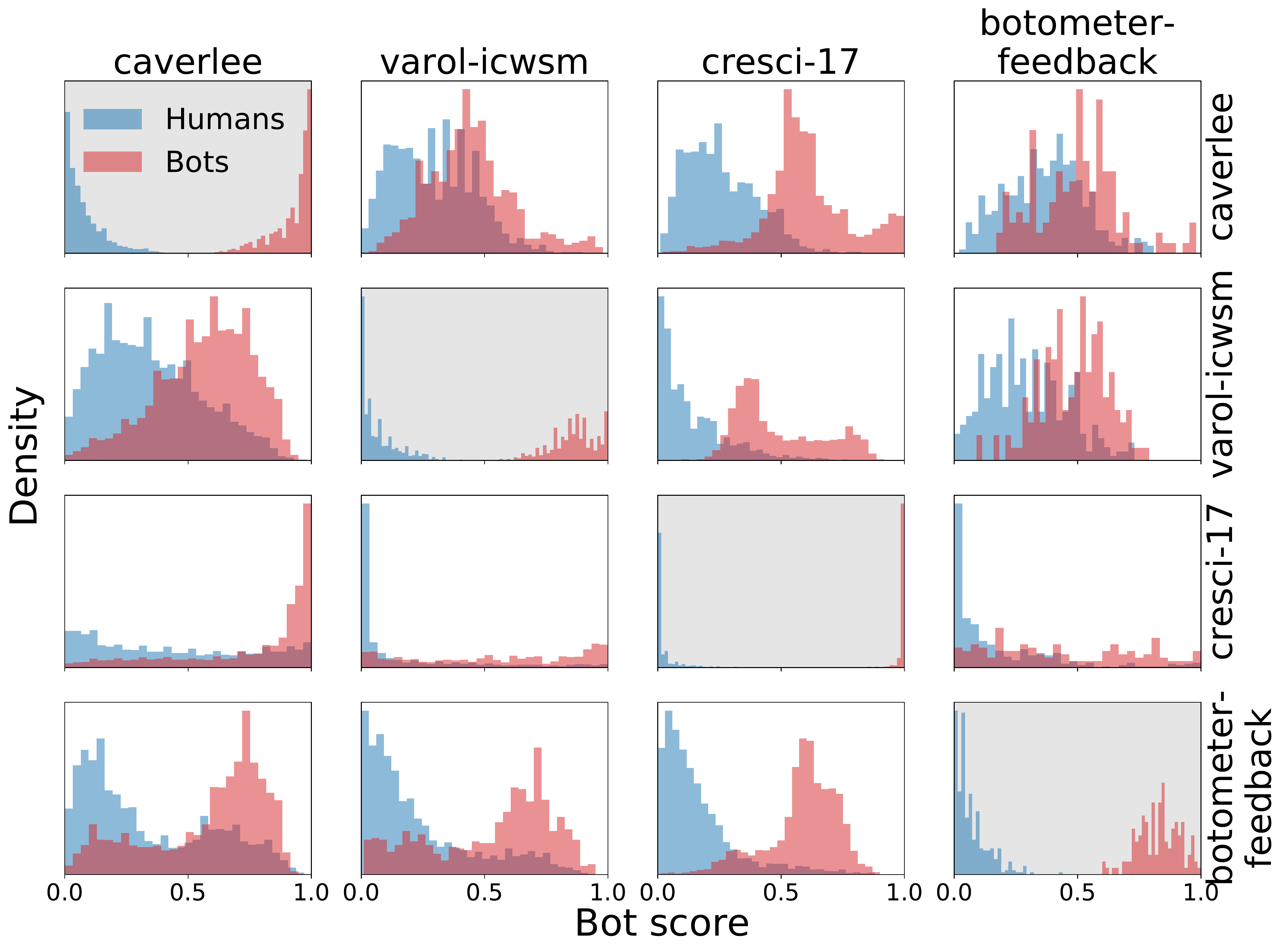}
    \caption{
    Bot score distributions for human (blue) and bot (red) accounts in each experiment.
    Models used in these experiments are trained on the datasets labeled along the rows and tested on the datasets listed in the columns.
    }
    \label{fig:densityPlotsMatrix}
\end{figure}

To interpret the cross-domain classification results, we plot the distributions of bot scores in Fig.~\ref{fig:densityPlotsMatrix}. A bot score is the output of a Random Forest classifier and corresponds to the proportion of decision trees in the ensemble that categorize the account as a bot. 
In the diagonal plots (in-domain tests), the density plots are left-skewed for humans and right-skewed for bots, representing a good separation and yielding high precision, recall, and F1.
For most of the cross-domain experiments, the score distributions are still left-skewed for humans, but not right-skewed for bots. 
This suggests that bot accounts tend to have lower cross-domain scores, resulting in lower recall. Human accounts, on the other hand, exhibit consistent characteristics across datasets. This observation suggests a way to improve generalization. 

\subsection{Predictability of different bot classes}

\begin{table*}[t!]
  \caption{Most informative features per bot class in \texttt{cresci-17}.}
  \label{tab:botclassFeatures}
  \footnotesize
  \centering
  \begin{tabular}{llll}
  \hline
  Rank & Traditional spambots & Social spambots & Fake followers  \\
    \hline
    1 & Std. deviation of adjective frequency & Tweet sentiment arousal entropy  & Max. friend-follower ratio \\
    2 & Mean follower count & Mean friend count & Std. deviation of tweet inter-event time \\ 
    3 & Tweet content word entropy & Mean adjective frequency & Mean follower count\\
    4 & Max. friend-follower ratio & Minimum favorite count & User tweet-retweet ratio \\
    5 & Max. number of retweet count & Tweet content word entropy &  Mean tweet sentiment happiness \\
    \hline
  \end{tabular}
\end{table*}

There are different kinds of bots. Consider the three different bot classes in the \texttt{cresci-17} dataset: traditional spambots, social spambots, and fake followers. We trained decision trees to discriminate each of these classes of bots from the others.  Table~\ref{tab:botclassFeatures} shows that different features are most informative for each class: traditional spambots generate a lot of content promoting products and can be detected by the frequent use of adjectives; social spambots tend to attack or support political candidates, therefore sentiment is an informative signal; finally, fake followers tend to have aggressive following patterns, flagged by the friend/follower ratio. 

\begin{figure}[t!]
    \centering
    \includegraphics[width=\columnwidth]{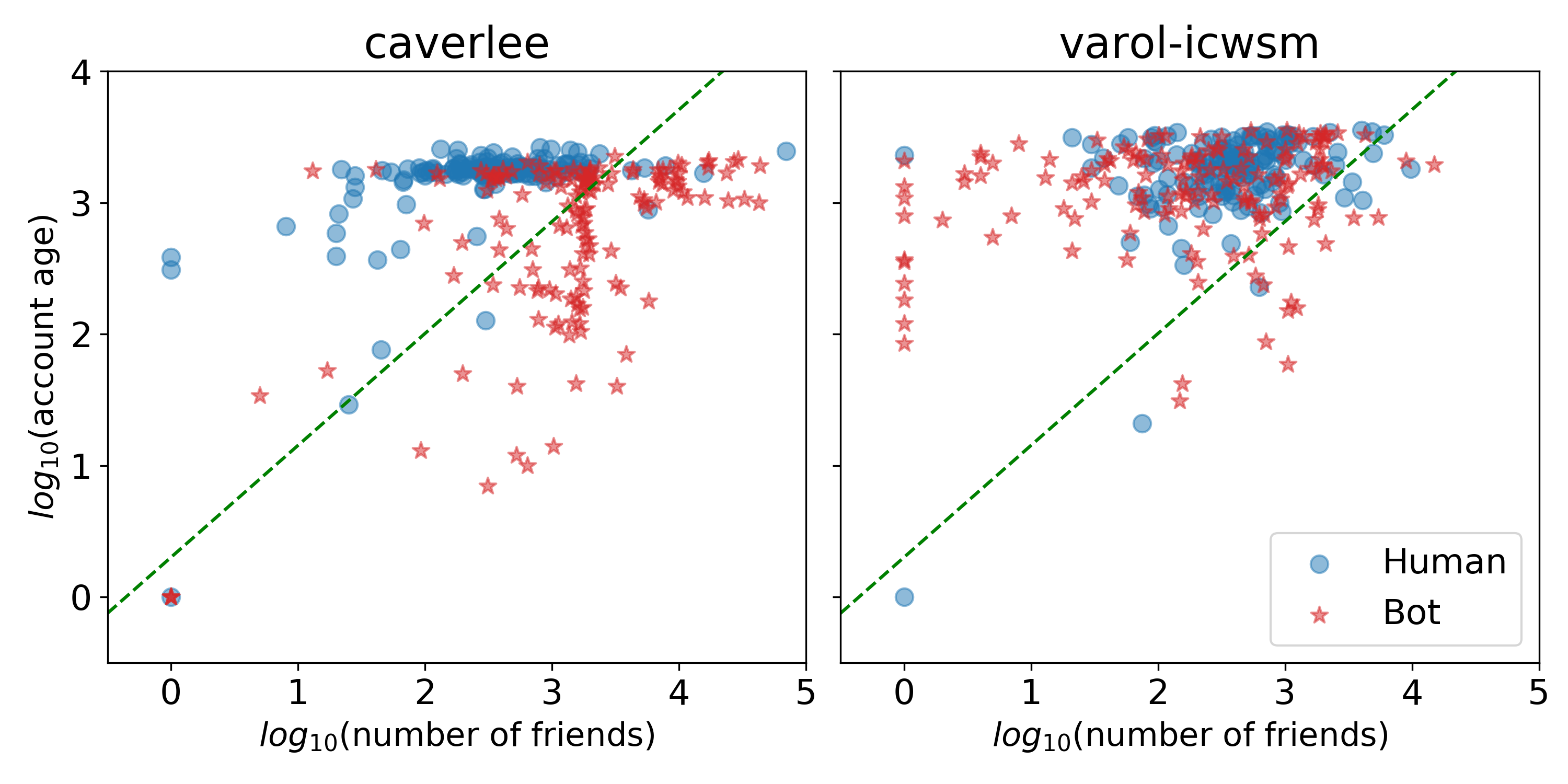}
    \caption{
    Separation of bots and humans based on two features in different datasets. Both plots show the logistic-regression decision boundary obtained from \texttt{caverlee}.
    }
    \label{fig:mixtureOfBotsHumans}
\end{figure}

Given such heterogeneity of bot behaviors, we conjecture that the drop in cross-domain recall can be attributed to the distinct discriminating features of accounts in different datasets. To explore this conjecture, let us use the Gini impurity score of a Random Forest classifier to find the two most informative features for the \texttt{caverlee} dataset, then train a logistic regression model on those two features. 
Fig.~\ref{fig:mixtureOfBotsHumans} visualizes bot and human accounts in \texttt{caverlee} and \texttt{varol-icwsm} on the plane defined by the two features.
For the in-domain case (left), the linear classifier is sufficient to separate human and bot accounts.
But in the cross-domain case (right), the same linear model fails, explaining the drop in recall: different features and distinct decision rules are needed to detect different classes of bots.  

\section{Methods}
                
\subsection{Baseline bot detection models}
\label{sec:baseline}

Before presenting the proposed method, let us select two baselines for evaluation. The first baseline is the current version of Botometer, often considered the state-of-the-art method for bot detection~\cite{yang2019arming}. The model is a Random Forest, a classification model that has proven to be effective in high-dimensional problems with complex decision boundaries. 
In this approach, we output the fraction of positive votes as a \emph{bot score}. 
Bots from all datasets are merged into a single bot (positive) class.
We refer to this baseline as \emph{Botometer-v3}.  
We also consider a variation of the Botometer baseline that does not consider a set of features describing time zone and language metadata, as those are no longer available through the Twitter API. We refer to this variation as \emph{Botometer-v3.1}. The two Botometer baselines use 1,209 and 1,160 features, respectively.

We use \emph{tweetbotornot2} as a second baseline. This model is based on a supervised classifier that considers over a hundred features in three main categories: user-level attributes, tweeting statistics, and text-based patterns. The motivation behind this choice of baseline is that \emph{tweetbotornot2} has been developed independently, is widely used, and is easily accessible by the general public via an R library (\url{tweetbotornot2.mikewk.com}).

\subsection{Proposed method}

\begin{figure}[t!]
    \centering
    \includegraphics[width=\columnwidth]{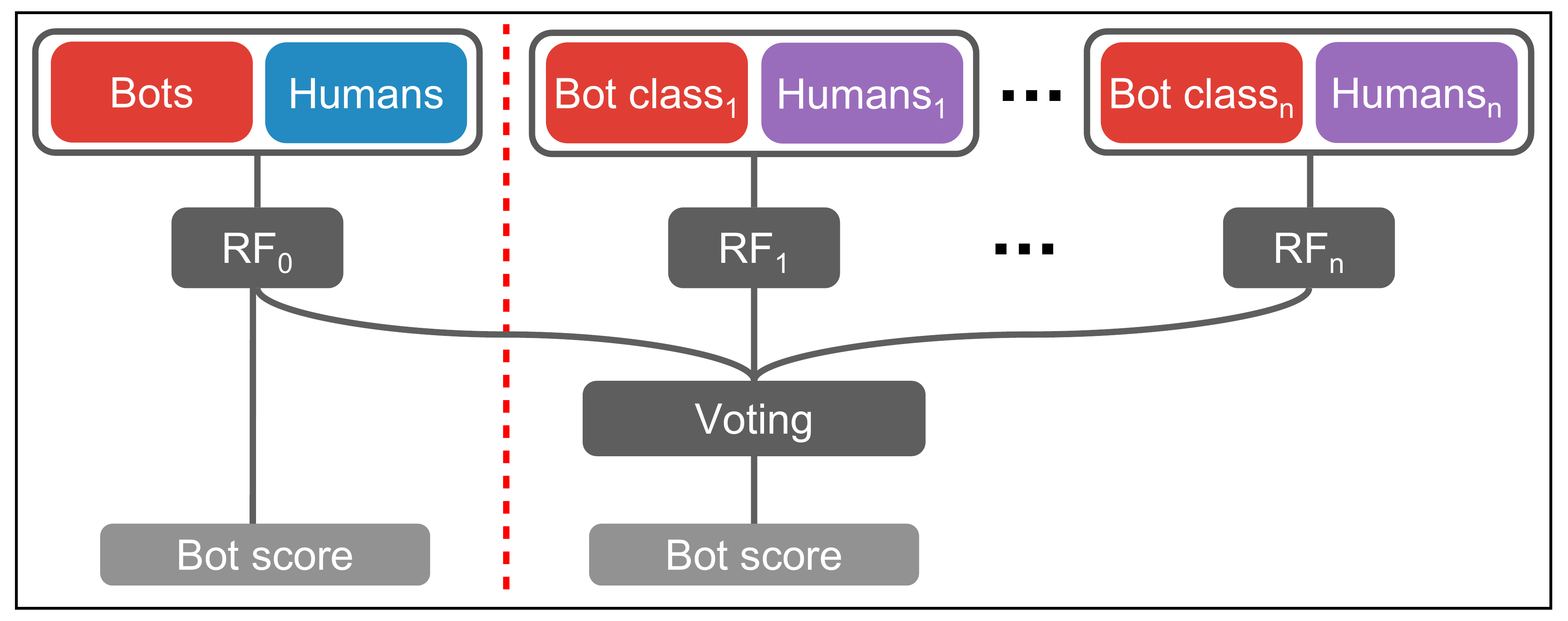}
    \caption{
    Illustration of the proposed model. The bot score from $RF_0$ corresponds to the previous version of Botometer, that from the voting module to the new (ESC) version.
    }
    \label{fig:methods_illustration}
\end{figure}

The proposed approach is inspired by the two empirical findings discussed in the previous section. 
First, inspired by the observation that human accounts are more homogeneous than bots across domains, we train a model on all human and bot examples across datasets and use it to identify likely humans. 
Second, since different bot classes have different sets of informative features, we propose to build specialized models for distinct bot classes. 
The specialized human and bot models are aggregated into an ensemble and their outputs are combined through a voting scheme.
We call this approach \emph{Ensemble of Specialized Classifiers} (ESC).

Fig.~\ref{fig:methods_illustration} illustrates the ESC architecture. The human detection subsystem actually corresponds to Botometer (the baseline classifier), and constitutes the left-most component $RF_0$ of the ensemble shown in the figure.
We then build specialized bot classifiers using Random Forest models ($RF_1 \dots RF_n$ in Fig.~\ref{fig:methods_illustration}). We use 100 decision tree estimators; all other parameters take the default values. 
Each specialized classifier $RF_i$ is trained on a balanced set of accounts from bot class $BC_i$ and an equal number $|BC_i|$ of human examples sampled from human accounts across all datasets.

A bot score is calculated by a voting scheme for the classifiers in the ensemble. Among the specialized bot classifiers, the one that outputs the highest bot score $s_i$ is most likely to have recognized a bot of the corresponding class. Therefore we use the maximum rule to aggregate the bot scores. For the human classifier $RF_0$, a low bot score $s_0$ is a strong signal of a human account. Therefore we determine the winning class as $i^* = \arg \max_i \{ s'_i \}$ where 
\[
s'_i = \left\{
 \begin{array}{ll}
 1-s_i  & \mbox{if } i = 0 \\
 s_i & \mbox{else.}
\end{array}
\right.
\] 
The ESC bot score is obtained by calibrating the score $s_{i^*}$ using Platt's scaling \cite{calibrationPaper}. 
As we see in \S~\ref{sec:results}, the maximum rule has the effect of shifting scores of likely bots toward one and the scores of likely humans toward zero.
Along with the bot score, ESC can also produce the bot class label $i^*$ as an explanatory outcome.

\section{Results}
\label{sec:results}

To train the specialized classifiers, we need coherent bot behaviors. For the experiments in this section, we organized the bot accounts in the training data into separate classes of bots: simple bots, spammers, fake followers, self-declared, political bots, and others.
Simple bots are derived from \texttt{caverlee}.
For spammer bots we use bot accounts in \texttt{pron-bots} and a subset of \texttt{cresci-17}. 
Fake followers include subsets of bot accounts in \texttt{cresci-17} and \texttt{vendor-purchased}. 
Self-declared bots are derived from \texttt{botwiki}. 
Political bots come from \texttt{political-bots} plus a subset of \texttt{astroturf}.
The rest of the bots captured in other datasets are aggregated into the ``others'' category. 

\subsection{In-domain performance}

\begin{figure}[t!]
    \centering
    \includegraphics[width=\columnwidth]{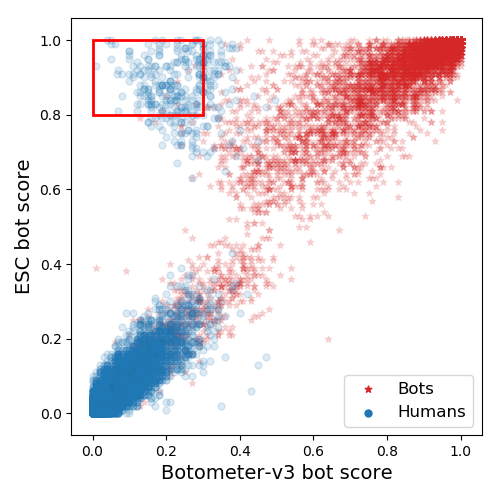}
    \caption{
    Correlation between bot scores obtained from the two methods.
    }
    \label{fig:botscoreIndomainScatter}
\end{figure}

Before discussing cross-domain performance, let us demonstrate that ESC is capable of detecting bots with good accuracy in the classic (in-domain) scenario. 
Using 5-fold cross-validation, ESC achieves an Area Under the ROC Curve (AUC) of 0.96, similar to the Botometer classifier (0.97 AUC). 
The scatter plot in Fig.~\ref{fig:botscoreIndomainScatter} shows a good agreement between the bot scores obtained with ESC and the \emph{Botometer-v3} baseline (Spearman's $\rho=0.87$). 

Let us pay special attention to  accounts having low Botometer and high ESC scores, in other words those that are likely human according to Botometer, but likely bots according to ESC (region highlighted in Fig.~\ref{fig:botscoreIndomainScatter}). These are the only cases where we observe a clear disagreement between the two methods on ground-truth labels. We focus on 332 accounts in this region that are labeled as human (75\% are from \texttt{caverlee}), which represent approximately 1\% of the examples labeled as human in the training data. 
One possible interpretation of the disagreement is that these accounts are incorrectly classified by ESC (false positives). Another possibility is that some of these accounts may have changed since they were manually labeled.  
Indeed, training datasets are subject to change over time as accounts become inactive, suspended, or get compromised by third-party applications. Such changes could lead to errors on ground-truth labels. 

Manual inspection of a random sample of 50 of the accounts highlighted in Fig.~\ref{fig:botscoreIndomainScatter} reveals that the human labels are no longer accurate for most of them --- they have been inactive for years, are currently devoted to spam diffusion, and/or are controlled by third-party applications. 
This suggests that ESC can identify impurities in the training data. While mislabeled accounts impair the performance of machine learning models, we conjecture that the ESC model is still able to recognize them because it is more robust to errors --- the incorrect labels only affect a subset of the classifiers.

\subsection{Cross-domain performance}
\label{sec:x-domain}

We want to demonstrate that the proposed ESC approach generalizes better to cross-domain accounts compared to the current version of Botometer. 
In this set of experiments, some datasets are held out in the training phase and are then used as cross-domain test cases: \texttt{cresci-stock}, \texttt{gilani-17}, \texttt{cresci-rtbust}, \texttt{kaiser-1}, \texttt{kaiser-2}, and \texttt{kaiser-3}. In addition,  \texttt{combined-test} combines these datasets while avoiding duplication.  
We focus on F1 and AUC as the accuracy metrics.  We obtain confidence intervals using bootstrapping with five samples of  80\% of the accounts in the test set.

\begin{figure}
    \centering
    \includegraphics[width=\columnwidth]{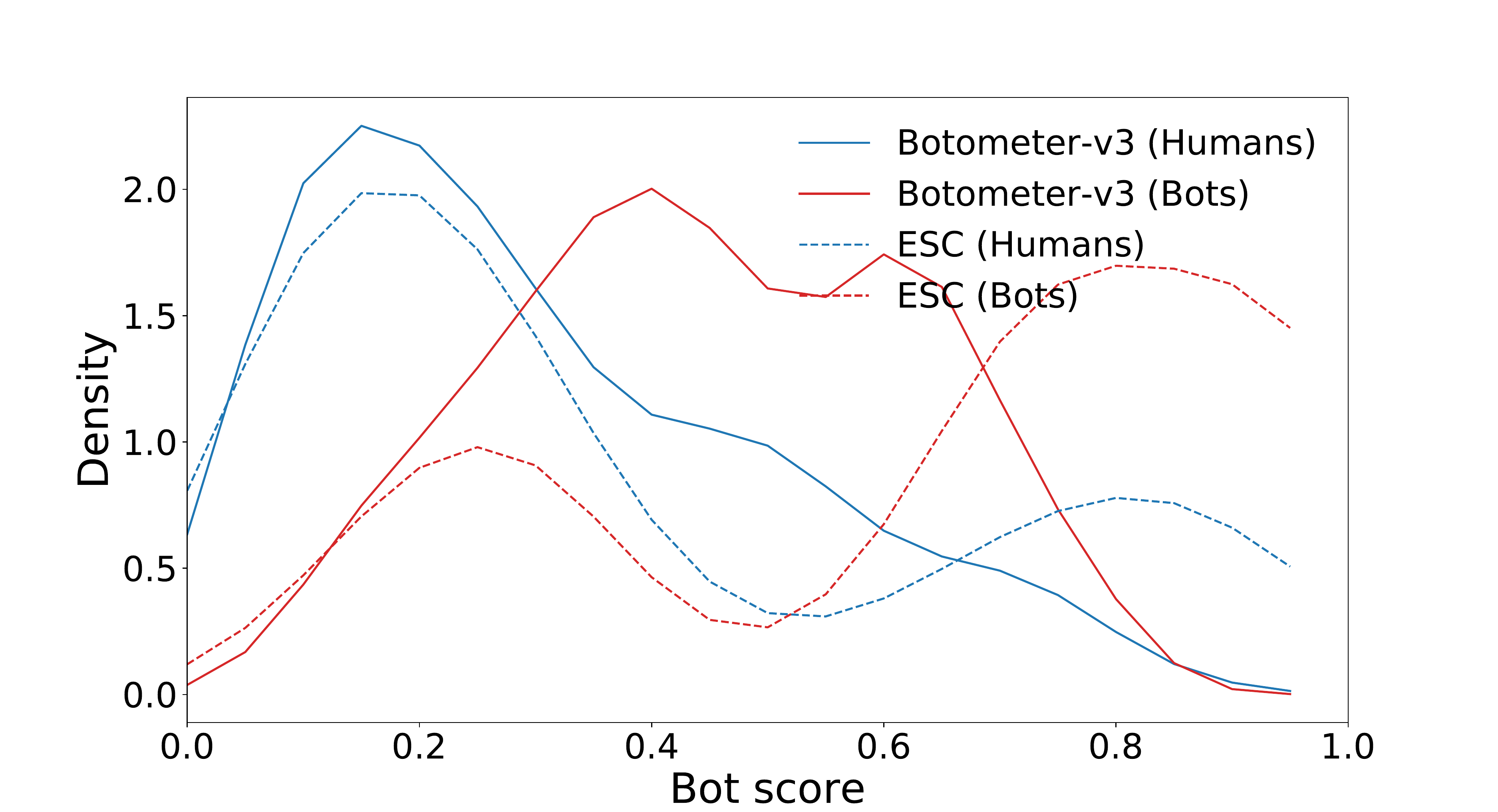}
    \caption{
    Distributions of bot scores (using KDE) for both methods on the hold-out \texttt{cresci-rtbust} dataset.
    }
    \label{fig:densityPlotsNewDomain}
\end{figure}

We compare the ESC approach with the \emph{Botometer-v3} baseline model~\cite{yang2019arming}, as well as the \emph{Botometer-v3.1} baseline to exclude the possibility that improvements are due to the removal of features based on deprecated metadata. The language agnostics version of Botometer (excluding English-based linguistic features) is used on the \texttt{kaiser-2} dataset because of its German tweet content.

To illustrate the main enhancement afforded by ESC, let us compare the distributions of bot scores generated by \emph{Botometer-v3} and ESC in a cross-domain experiment. Due to space limitations, we illustrate in Fig.~\ref{fig:densityPlotsNewDomain} just one case where \texttt{cresci-rtbust} is used as test set; other cases are similar. Both methods tend to yield low scores for human accounts, as the same classifier ($RF_0$) is used. 
On the other hand, ESC produces significantly higher scores than \emph{Botometer-v3} on bot accounts.
This is a result of the maximum rule and leads to higher cross-domain recall, or better generalization. 

\begin{figure}
    \centering
    \includegraphics[width=\columnwidth]{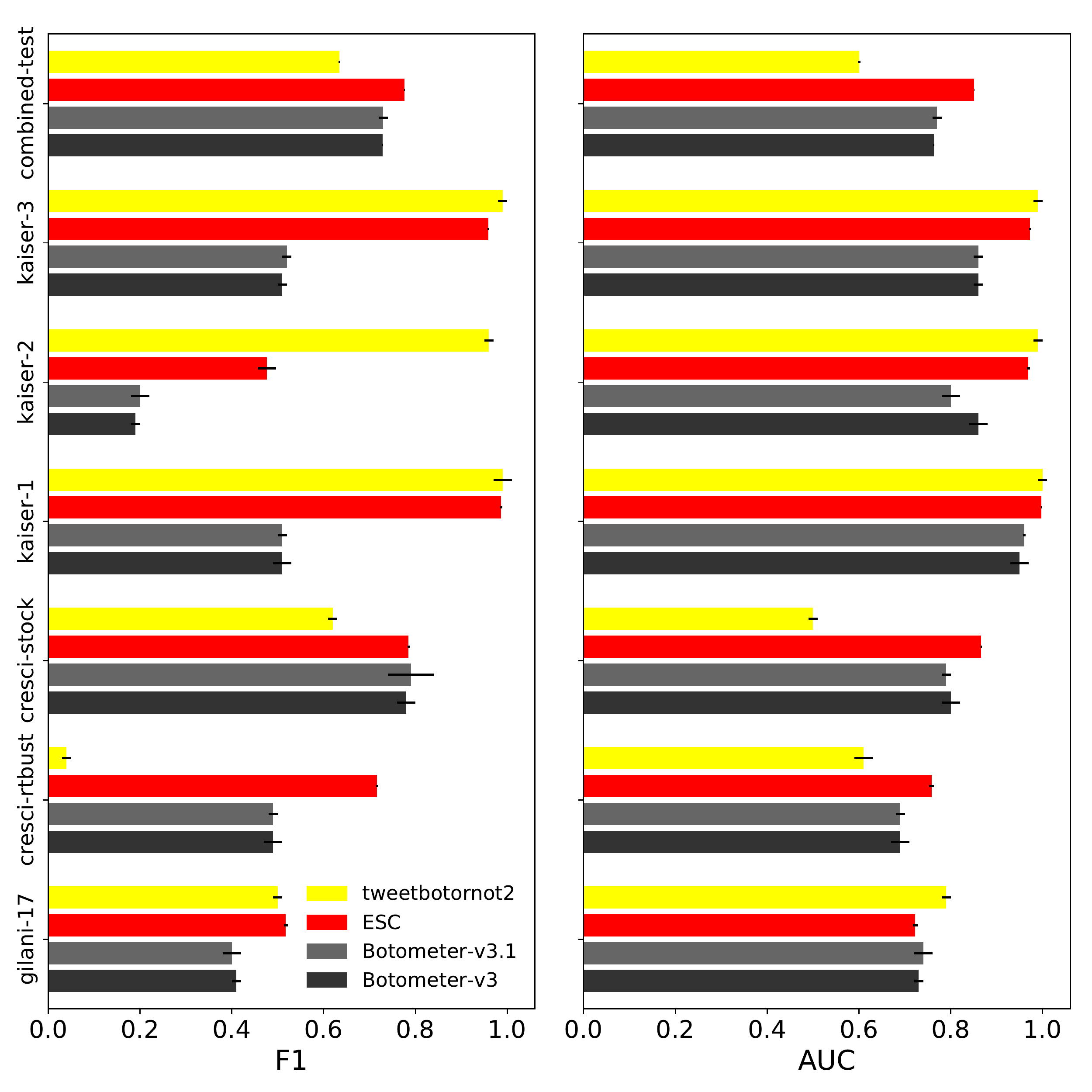}
    \caption{
    F1 (left) and AUC (right) of ESC and baseline methods on hold-out test datasets.
    Error bars indicate 95\% confidence intervals. 
    Note that \texttt{kaiser-1} and \texttt{kaiser-3} are not `hold-out' as there is overlap with the training sets (see text).
    }
    \label{fig:f1ErrorBar}
\end{figure}

Fig.~\ref{fig:f1ErrorBar} shows that ESC outperforms the Botometer baseline in most cases. 
On average across the six datasets, recall goes from 42\% to 84\% (an improvement of 100\%) while precision increases from 52\% to 64\%. As a result, F1 increases from 47\% to 73\% (an improvement of 56\%). 
On the \texttt{combined-test} dataset, recall goes from 77\% to 86\%, precision stays at 70\%, and F1 goes from 73\% to 77\% (an improvement of 5\%).
Comparisons based on AUC scores are similar. 

The \texttt{kaiser-1} and \texttt{kaiser-3} datasets include bots from \texttt{botwiki}, which are part of the ESC training data. Therefore these two cannot be considered completely hold-out datasets, but are included nonetheless because they were used to highlight weaknesses of \emph{Botometer-v3} in a recent independent report~\cite{botometerCriticismHarvard}, so they provide us with an opportunity to demonstrate the performance of the latest Botometer model.  
Even if we exclude \texttt{botwiki} from the training data, ESC still outperforms \emph{Botometer-v3}. For example, it yields an F1 score of 0.84 on \texttt{kaiser-1} and 0.80 on \texttt{kaiser-3}.

ESC yields F1 better than or comparable with \emph{tweetbotornot2} on all datasets except those from \citet{botometerCriticismHarvard}. The AUC metric is comparable on those datasets, while \emph{tweetbotornot2} wins on \texttt{gilani-17} and ESC wins on the other datasets. 
On the \texttt{combined-test} dataset, ESC outperforms \emph{tweetbotornot2} on both F1 and AUC. 

In interpreting these results,  note that \texttt{kaiser-3} contains human accounts from \texttt{kaiser-2} and bot accounts from \texttt{kaiser-1}, so they are not independent.
\texttt{kaiser-2} includes only 27 bot accounts; the F1 score is sensitive to this class imbalance. 
ESC is comparable to \emph{tweetbotornot2} on this dataset when using the AUC metric, which is not sensitive to class imbalance. 
Furthermore, while we do not know how the \emph{tweetbotornot2} baseline was trained, it uses a feature for ``verified'' profiles and tends to assign a low bot score to them, even automated ones such as \texttt{@twittersupport}. 98\% and 72\% of accounts labeled human in \text{kaiser-1} and \texttt{kaiser-3} respectively are verified. This biases the results in favor of \emph{tweetbotornot2}.

\subsection{Model adaptation}

Real-world applications of social bot detection always face the challenge of recognizing novel behaviors that are not represented in the training data. 
Periodic retraining to include newly annotated datasets helps systems adapt to these unseen examples.
A common approach is to train a new classifier from scratch including both old and new training data, which may not be efficient. The proposed ESC method alleviates this problem because we can add a new classifier $RF_{n+1}$ to the ensemble to be trained with the new data, without retraining the existing classifiers.

Let analyze how quickly the \emph{Botometer-v3} and ESC models adapt to a new domain. 
To quantify this, we split the data from a hold-out domain into two random subsets for training and testing. 
Results are presented using  \texttt{varol-icwsm} as the hold-out domain for both models. (We reach similar conclusions using other hold-out datasets.) 800 examples from the hold-out dataset are used for training.  In each iteration we randomly sample 50 examples and add them to the training set. In the \emph{Botometer-v3} case, we retrain the entire classifier ($RF_0$), whereas in the ESC case we only train a newly added specialized classifier ($RF_{n+1}$).

\begin{figure}[t!]
    \centering
    \includegraphics[width=\columnwidth]{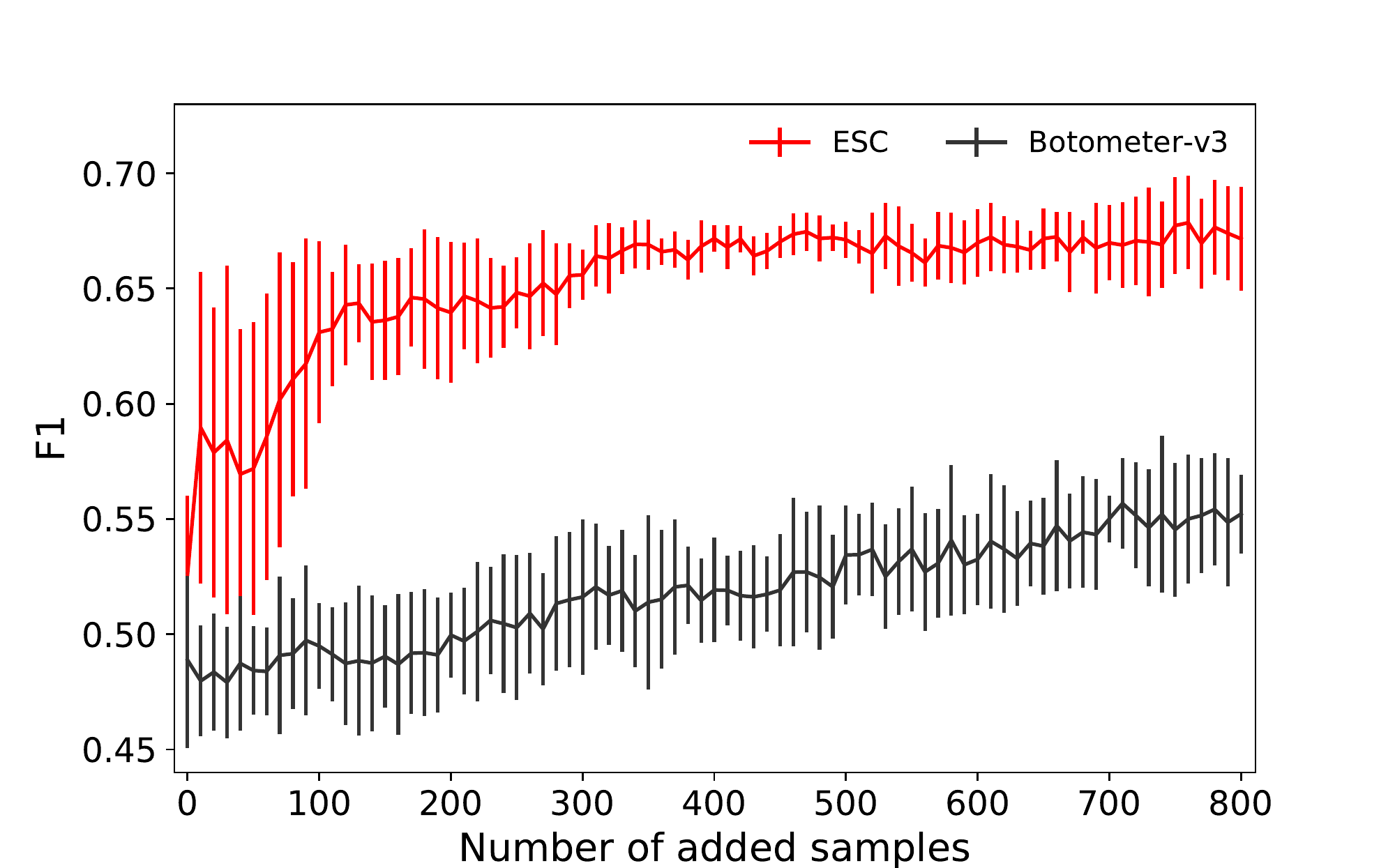}
    \caption{Adaptation of two methods to new examples from the \texttt{varol-icwsm} dataset, which was held out from the training data.}
    \label{fig:modeladaptationResults}
\end{figure}

Fig.~\ref{fig:modeladaptationResults} shows how the F1 score on the test data improves as a function of the number of training examples from the hold-out domain.
The \emph{Botometer-v3} baseline adapts more slowly. We can interpret this result by recalling that the size of the hold-out dataset is small compared to the training size (over 67,000 examples in \emph{Botometer-v3}).  The decision trees use Gini gain as a feature selection criterion, therefore the number of examples sharing the same informative features affects the selection of those features. As a result, the old bot classes with more examples dominate and the classifier struggles to learn about the new domain. On the other hand, the ESC architecture quickly learns about new bots through the new classifier, which starts from scratch, while what was learned about the old ones is preserved in the existing classifiers. 
This means that fewer labeled examples are needed to train a new specialized classifier when novel types of bots are observed in the wild.

\section{Related work}

Different approaches have been proposed for automatic social bot detection. Crowdsourcing is convenient and one of the first proposals to collect annotated data effectively \cite{wangCrowdSourcedBotdetection}, but annotation has limitations due to scalability and user privacy. Thus automatic methods are of greater interest, especially for social media services that deal with millions of accounts. The structure of the social graph captures valuable connectivity information. Facebook employed an immune system \cite{immuneFb} that relied on the assumption that sybil accounts tend to connect mostly to other sybil accounts while having a small number of links to legitimate accounts.

Supervised machine learning approaches, such as the one proposed in this paper, extract various features from an account's profile, social network, and content \cite{miller2014twitter,subrahmanian2016darpa,davis2016botornot,varol2017online,gilani2017classification,yang2019arming,yang2020scalable}. 
These methods rely on annotated datasets for learning the difference between bot and human accounts. 
However, since bots change and evolve continuously to evade detection, supervised learning algorithms need to adapt to new classes of bots~\cite{ferrara2016rise,varol2018deception,yang2019arming}. 

Some unsupervised learning methods have been proposed in the literature \cite{cresci2017paradigm,unsupHoloscope}. They can be less vulnerable to performance decay across domains. They are especially suitable for finding coordination among bots \cite{ahmed2013generic,miller2014twitter,pacheco2020uncovering}. Since accounts in a coordinated botnet may not appear suspicious when considered individually, supervised methods would miss them \cite{cresci2017paradigm,grimme2018changing}. Identifying botnets requires analysis of the activity of multiple accounts to reveal their coordination. Depending on the type of bots, similarity can be detected through tweet content \cite{chen2018unsupervised,kudugunta2018deep}, temporal features in the timelines \cite{chavoshi2016debot,cresci2016dna}, or  retweeting behavior \cite{vo2017revealing,mazza2019rtbust}.

A recent research direction is to address the limits of current bot detection frameworks in an adversarial setting. 
\citet{cresci2018reaction} predict that  future techniques will be able to anticipate the ever-evolving spambots rather than taking countermeasures only after seeing them. 
The performance decay of current detection systems in the wild was reported by \citet{cresci2017paradigm}, who showed that Twitter, human annotators, and state-of-the-art bot detection systems failed at discriminating between some new social spambots and genuine accounts. 
In agreement with the present findings, \citet{echeverria2017discovery2} suggest that detecting different types of bots requires different types of features.
\citet{de2018lobo} proposed the leave-one-botnet-out methodology, to highlight how detection methods do not generalize well to unseen bot classes due to bias in the training data. Even a classifier trained on 1.5 million accounts and 22 classes of bots is incapable of identifying new classes of bots. Here we follow the leave-one-class-out methodology to evaluate the generalization power of the proposed approach.

Some papers have characterized bot classes \cite{botTypes2014}. \citet{lee2011seven} define traditional spammers, mention spammers, friend infiltrators, and social spammers in their dataset. \citet{cresci2017paradigm} highlight that it is hard for human annotators to assign one label to one account to describe its bot class. They also report that some bot classes, like social spambots, are hard to distinguish from human accounts. 
Despite the existence of different types of bots, no systems have been presented in the literature to automatically identify the type of a bot as the method proposed here.

\citet{botometerCriticismHarvard} have criticised Botometer (more specifically \emph{Botometer-v3}) for its high false positive and false negative rates. As we have discussed in this paper, we share these concerns and acknowledge that like any supervised learning model, Botometer makes mistakes. Indeed, the new version introduced in the paper is motivated by this issue and partly addresses it by improving cross-domain recall (false negatives). 
At the same time, \citet{botometerCriticismHarvard} may overestimate the false positive rate by assuming that no politician account uses automation. We believe this assumption is not realistic, considering these accounts are often managed by media teams and use scheduling tools for content creation. In fact, Botometer currently does not use the ``verified'' status as a feature because it could lead to false negatives~\cite{varol2019journalists}. Another source of bias is that \citet{botometerCriticismHarvard} overlook accounts that are no longer available due to suspension, possibly leading to an underestimation of both precision and recall.

\section{Botometer-v4 Deployment}

\begin{figure}
    \centering
    \includegraphics[width=\columnwidth]{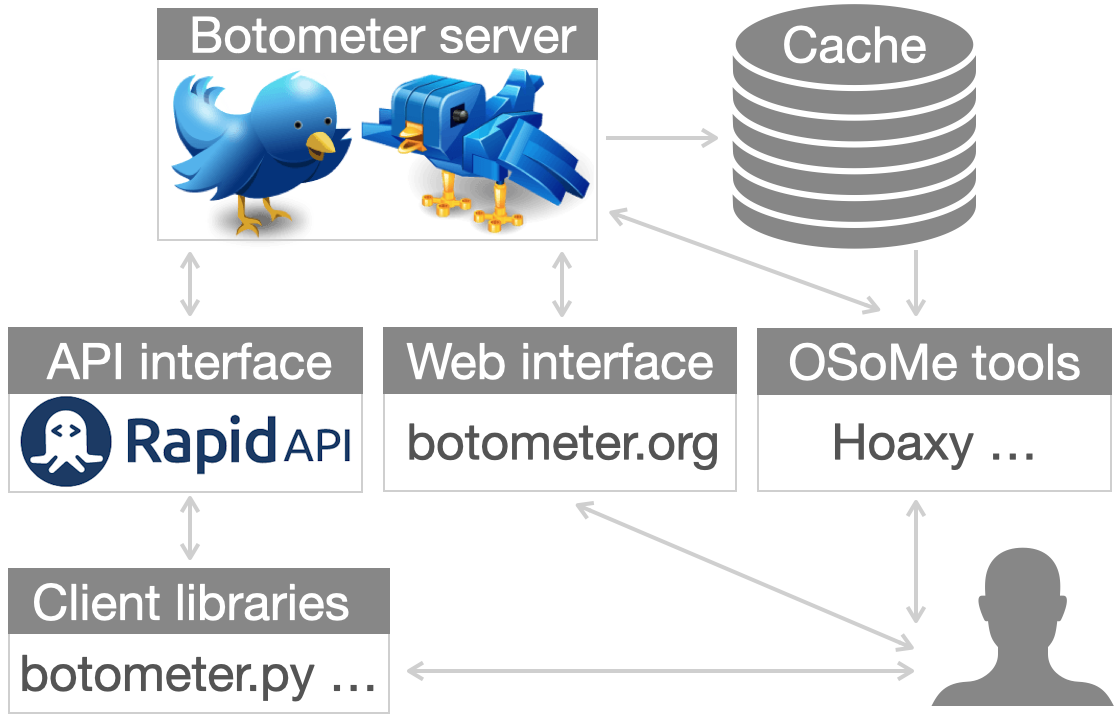}
    \caption{
    The architecture of the Botometer ecosystem.
    }
    \label{fig:botometer_architecture}
\end{figure}

In light of these results, we are deploying ESC in the newest version of Botometer (v4), a tool to detect social bots in the wild that is available through the Observatory on Social Media (OSoMe) at Indiana University. Botometer can be accessed both through an interactive website (\url{botometer.org}) and programmatically through a public API (\url{rapidapi.com/OSoMe/api/botometer-pro}). Fig.~\ref{fig:botometer_architecture} illustrates the system architecture. 

\begin{figure}
    \centering
    \includegraphics[width=\columnwidth]{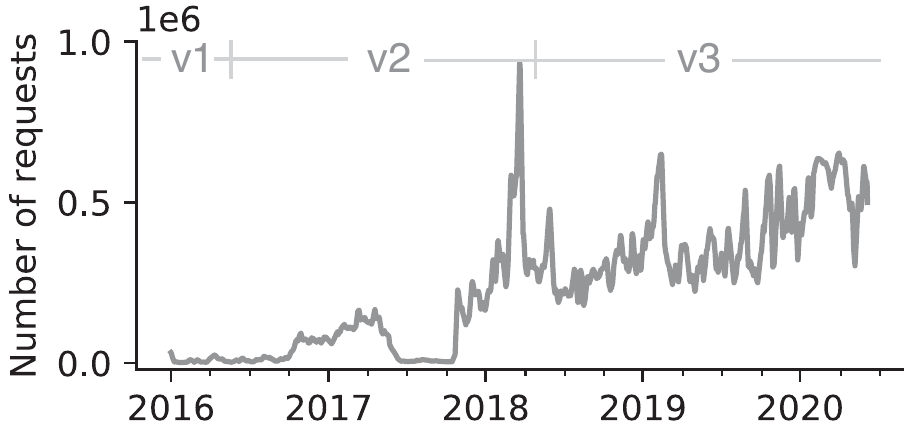}
    \caption{
    Daily requests of the Botometer API.
    Annotations indicate the versions of the models deployed in different time periods.
    }
    \label{fig:dailyRequests}
\end{figure}

The deployed system is implemented in Python with the MKL library for efficiency. Random Forest classifiers are implemented using the \textit{scikit-learn} package~\cite{pedregosa2011scikit}. Since ESC provides a good correspondence between scores and binary labels of human and bot accounts, no calibration is applied.  
We performed load tests submitting queries from 20 simultaneous jobs running on a supercomputer, yielding an average response time below 100ms per query. 
This should provide ample scalability, considering the popularity of the tool among researchers and practitioners --- at the time of this writing, it fields over half a million queries daily (Fig.~\ref{fig:dailyRequests}).

To train the specialized classifiers of the deployed model on homogeneous bot types, we rearranged the datasets in Table~\ref{table:dataset}. The classes are similar to those described in \S~\ref{sec:results}, with several modifications: (i)~we removed \texttt{caverlee} based on the analysis in Fig.~\ref{fig:botscoreIndomainScatter}; (ii)~we used the full \texttt{astroturf} dataset; (iii)~we added a financial bot class based on \texttt{cresci-stock}; (iv)~we extended the class of other bots using \texttt{cresci-rtbust}, \texttt{gilani-17}, the new bots in \texttt{kaiser-1}, and the German bots in \texttt{kaiser-2}; and (v)~we added human accounts from the \texttt{combined-test} dataset and \texttt{midterm-2018}. 
The final model yields an AUC of 0.99.

The new front-end of the Botometer website and the Botometer API report the scores obtained from the six specialized classifiers (fake followers, spammers, self-declared, astroturf, financial, others). In this way, the tool offers greater transparency about the decision process by allowing inspection of the outputs of different ensemble components and by providing interpretable class labels in output.

\section{Conclusion}

The dynamic nature of social media creates challenges for machine learning systems making inferences and predictions on online data. On the one hand, platforms can change features, require models to be retrained. Further difficulties arise as accounts used for training change behavior, become inactive, compromised, or are removed from the platform, invalidating ground-truth data. 
On the other hand, account behaviors can change and evolve. As is typical in adversarial settings, automated accounts become more sophisticated to evade detection. The emergence of more advanced bot capabilities brings additional challenges for existing systems that struggle to generalize to novel behaviors.

Despite impressive results when training and test sets are from the same domain --- even using cross-validation --- supervised models will miss new classes of bots, leading to low recall. 
We demonstrate in this paper that the performance deterioration observed for out-of-domain accounts is due to heterogeneous bot behaviors that require different informative subsets of features.
Inspired by this, we presented a novel approach for the automatic identification of novel social bots through an ensemble of specialized classifiers trained on different bot classes. 
We demonstrated empirically that our proposed approach generalizes better than a monolithic classifier and is more robust to mislabeled training examples. However, our experiments show that cross-domain performance is highly sensitive to the datasets used in training and testing supervised models; it is easy to cherry-pick examples that make any given method look bad.

The proposed architecture is highly modular as each specialized classifier works independently, so one can substitute any part with different models as needed. 
We can also include additional specialized classifiers when new annotated datasets become available.
We showed that this approach allows the system to learn about new domains in an efficient way, in the sense that fewer annotated examples are necessary.

In future work, we would like to investigate methods to recognize the appearance of a new type of bots that warrants the addition of a new classifier to the ensemble.
We could also design an unsupervised method to cluster similar accounts across datasets automatically, and assign homogeneous accounts to each of the specialized classifiers.
Finally, one could design active learning query strategies to make the retraining process even more efficient than with random selection. This would be useful when reliable user feedback is available to be used as an oracle.

\section{Acknowledgments}

We are grateful to Clayton A. Davis and Emilio Ferrara who contributed to early versions of Botometer, and Chris Torres-Lugo for helping with the \texttt{astroturf} dataset.
This work was supported in part by DARPA (grant W911NF-17-C-0094), Knight Foundation, and Craig Newmark Philanthropies. 

\newpage





\end{document}